\newcommand{\CLASSINPUTtoptextmargin}{0.75in}
\newcommand{\CLASSINPUTbottomtextmargin}{1in}

\documentclass[conference]{IEEEtran}
\IEEEoverridecommandlockouts

\usepackage[english]{babel}
\usepackage{cite}
\usepackage{amsmath,amssymb,amsfonts}
\usepackage{algorithmic}
\usepackage{graphicx}
\usepackage{textcomp}
\usepackage{xcolor}
\usepackage{hyperref}
\usepackage{comment}   
\usepackage{tikz}

\includecomment{review}
\renewcommand{\comment}[2]{}
\begin{review}
\renewcommand{\comment}[2]{\medskip\noindent\textcolor{#1}{\footnotesize #2}\medskip}
\end{review}

\newcommand{\eqdef}{\triangleq}

\newcommand\acceptedtext{%
	\footnotesize This article has been accepted for publication in proceedings of the 2022 IEEE 9th International Workshop on Metrology for AeroSpace (MetroAeroSpace), but has not been fully edited. Content may change prior to final publication. \\
	Citation information: DOI 10.1109/MetroAeroSpace54187.2022.9856398.}
\newcommand\acceptednotice{%
	\begin{tikzpicture}[remember picture,overlay]
		\node[anchor=north,yshift=-6pt] at (current page.north) {%
			\begin{minipage}{\textwidth}
				\center \acceptedtext
		\end{minipage}};
	\end{tikzpicture}%
}

\begin{document}

\title{A cloud-assisted ADS-B network for UAVs \\ based on SDR
\thanks{The work of I.~Iudice and D.~Pascarella was partially
supported by the CIRA project ``MATIM''.}
}

\author{\IEEEauthorblockN{Giacinto Gelli}
\IEEEauthorblockA{\textit{Dipartimento di Ingegneria Elettrica} \\
\textit{e Tecnologie dell'Informazione} \\
\textit{Università Federico II di Napoli}\\
Napoli I-80125, Italy \\
\href{mailto:gelli@unina.it}{gelli@unina.it}}
\and
\IEEEauthorblockN{Ivan Iudice, Domenico Pascarella}
\IEEEauthorblockA{\textit{Department of Reliability and Security} \\
\textit{Italian Aerospace Research Centre}\\
Capua I-81043, Italy \\
{[\href{mailto:i.iudice@cira.it}{i.iudice},
\href{mailto:d.pascarella@cira.it}{d.pascarella}]}@cira.it}
}

\maketitle
\acceptednotice

\thispagestyle{plain}
\pagestyle{plain}

\begin{abstract}
Integration of Unmanned Aerial
Vehicles (UAVs) or ``drones'' into the
civil aviation airspace is a problem of increasing
interest in the aviation community,
as testified by many initiatives
developed worldwide.
Many traditional surveillance solutions
for manned aircrafts employ the
Automatic Dependent System-Broadcast (ADS-B) technology,
which however might present
several drawbacks when used for UAVs,
especially smaller ones
and/or those flying at very low altitudes.
We present in this paper
a cloud-based surveillance
solution for UAVs, which can be considered
as an enhancement of a conventional ADS-B system.
The proposed solution leverages inexpensive
on-board transceivers for transmitting
positional messages from the UAVs to the ground.
A network of ADS-B gateways,
based on
the software-defined radio (SDR)
paradigm, format the positional messages
into valid ADS-B signals and rebroadcast
them in the air, allowing thus to emulate
a true ADS-B system and overcoming
the main disadvantages
of the conventional implementation.
A preliminary performance
analysis of the proposed approach,
based on queuing theory, shows
the main tradeoffs of the
considered approach.
Moreover, a physical-layer laboratory implementation
of the proposed solution is presented,
based on off-the-shelf SDR hardware,
which is programmed using the open-source
GNU Radio environment.

\end{abstract}

\smallskip


\section{Introduction}

An Unmanned Aerial Vehicle (UAV),
shortly known as ``drone'',
is an aircraft with no human pilot on-board.
It represents the central element
of an Unmanned Aerial System (UAS),
which is the set of the aircraft and all the
other elements supporting its service.
Without the need of an on-board pilot,
drones were originally designed
to accomplish military
tasks \cite{Lopez2008}.
However, recent advances in drones' technology have
allowed the emergence of a wide new range of applications
in the civil domain, as highlighted by the
Single European Skies ATM Research Joint Undertaking (SESAR JU)
in its outlook study \cite{SESARJU2016}.
According to such a study, the role of drones is likely to
expand up to 2050 in many civil sectors, including
agriculture, energy, public safety and security,
e-commerce and delivery,
mobility and transport.

In view of the growing demand
for civil drone services
and their impacts in terms of economic growth
and societal benefits,
a key problem is to extend traditional
Air Traffic Management (ATM) systems
to implement \textit{Umanned Aircraft System Traffic Management}
(UTM) systems, aimed at safely and efficiently
managing small UAVs flying in low airspace.
To this aim, the SESAR JU started the
\textit{U-space programme} in 2016, defined
\cite{SESARJU2017} as
a set of services and specific procedures
designed to support safe, efficient and secure access to airspace
for large numbers of drones.
A similar evolutionary
pathway is being followed
in USA, where UTM is under development in the
\textit{Next Generation Air
Transportation System}
(NextGen) programme of the
Federal Aviation Administration
(FAA) \cite{FAA2020}.

To obtain
the required data for flight
situational awareness,
UTM systems shall rely upon a surveillance
infrastructure, which considers both the
unmanned traffic and
its interaction with manned traffic.
Several initiatives have proposed
the application of
\textit{Automatic Dependent Surveillance-Broadcast} (ADS-B)
for the safe integration of drones in the civil airspace.
The current ADS-B network infrastructure is composed by
cooperative aircrafts periodically broadcasting
their own positional
data%
\footnote{Positioning is typically achieved by collecting data
from different sensors, such as GPS, ultrasound, LIDARs, cameras, IMUs,
and performing sensor fusion by means of Kalman filters \cite{Minucci2020}.}
through Mode-S Extended Squitter ($1090$ MHz) \textit{ADS-B Out} messages
\cite{sun1090mhz},
and receiving nodes called \textit{ADS-B In} using
such information for ATM operations
and/or providing global flight tracking services, such
as, e.g., Flightradar24%
\footnote{\url{https://www.flightradar24.com}.}.
Each aircraft may in turn
be equipped with an ADS-B In receiver
to enhance its on-board
situational awareness
and to readily engage separation
maneuvers when needed.

ADS-B is already approved for use in
civil ATM and represents
a cost-effective surveillance technology with
great potential for novel applications, such as UTM.
For example, the European CORUS (Concept of Operations
for European Unmanned Traffic Management Systems) project
has defined a U-space Concept of Operations (ConOps), wherein ADS-B is
highlighted as a potential surveillance technology for the
\textit{electronic conspicuity}, i.e.,
a capability that enables the
broadcast or relay of an ownership’s location 
or position
to other airspace users and ground operators \cite{CORUS2019}.
Moreover, in their white papers about UTM,
Amazon \cite{Amazon2015}
and Google \cite{Google2015} have proposed ADS-B
as an essential asset for safe and cooperative
integration of manned/unmanned traffic.

However, several problems have
been identified when using standard ADS-B
for UTM.
For example, the analysis in \cite{Doole2018}
indicates a lack of
real-time position information at
the Very Low Level (VLL) altitude (under $500$ feets),
due to the current
number of on-board ADS-B
receivers and transponders,
which is insufficient for
the surveillance of high-density
drone operations in an
urban airspace.
To solve this problem, it is proposed
in \cite{Doole2018}
to increase
the density of ADS-B receivers
for capturing position information,
by mandating all aircrafts
to employ ADS-B transponders.

Instead, in \cite{Strohmeier2014} it is argued
that the integration of ADS-B as a core part
of the future ATM/UTM systems may be compromised
by issues related to severe message losses
(caused by the growing traffic on the channel)
and open security concerns,
due to the cheap and
easy availability of Software-Defined Radio (SDR) devices,
which can be easily employed to sniff, jam
or spoof legitimate ADS-B traffic.
Another issue is represented by the power consumption,
since standard ADS-B Out transponders use $200$ W transmit power,
whereas small transponders may consume up to $20$ W,
which is still too high for a small
battery-powered drone \cite{Minucci2020}.
Lastly, specialized ADS-B transceivers for drones are
much more expensive than drone themselves due to
the certification processes, and are affordable
only for bigger drones \cite{Minucci2020}.

In this paper,
to solve some of the previously proposed
problems, we propose an innovative
cloud-based surveillance
solution for UAVs,
which is a simple add-on
to the conventional ADS-B system.
The proposed solution leverages inexpensive
on-board transceivers for transmitting
positional messages from UAVs to the ground.
A network of ADS-B gateways based on the SDR paradigm,
after formatting the positional messages into valid ADS-B signals,
rebroadcasts them in the air, which allows one to emulate
a true ADS-B system.
The proposed solution employs SDR techniques
at the gateways,
which is a convenient tool for its
fast prototyping and flexibility features.
In particular, our SDR implementation is
based on the open-source
GNU Radio framework%
\footnote{\url{http://www.gnuradio.org}.}.
A preliminary performance
analysis, in terms of
overall system capacity and latency,
is also provided, based on queueing theory,
under some simplifying
assumptions.



\subsection{Related work}

Different research projects are facing the challenge
of designing ADS-B-based solutions for
the surveillance service in UTM.
In general, the addressed solutions
represent low-power ADS-B variants.
Indeed, reference \cite{Guterres2017} provides an analysis
of the impact on ADS-B performance from a shared-use operation by drones.
The analysis indicates that the key parameters
are drones' ADS-B transmission power and traffic density,
which should be balanced to attain an acceptable
demand on the ADS-B channel
in high-density traffic areas.

Within its UTM project, the National Aeronautics and
Space Administration (NASA) has been analyzing
the application of ADS-B for cooperative
surveillance of drones since 2015 \cite{Lozano2015}.
For example, one of their research works has provided
detailed simulation results of the ADS-B technology
and of the related \textit{detect and avoid }(DAA)
algorithms in mixed large-density
manned/unmanned environments
\cite{Matheou2018}.
Other NASA's works regard the Integrated Configurable Algorithms
for Reliable Operations of Unmanned Systems (ICAROUS)
software architecture, that is, a set of highly-assured algorithms
for building safety-centric and autonomous application
of unmanned aircrafts in UTM systems \cite{Consiglio2016}.
Several flight tests have been performed for the
ICAROUS Sense and Avoid Characterization (ISAAC)
to evaluate \cite{Duffy2019}: (i) the effectiveness of ADS-B
receivers for drones to receive position reports for ICAROUS
as a source of cooperative traffic surveillance;
(ii) the use of ADS-B for drones.
These tests have validated low-powered ADS-B transmissions ($0.4$ W)
as an option for drone-to-drone applications.

Moreover, private companies are working on ADS-B variants for drones.
For example, nearly all the new drones released by Da-Jiang Innovations (DJI)
will have the AirSense feature, which is an alert system to
give drone pilots an enhanced situational awareness about
manned aircraft by means of an ADS-B In component.%
\footnote{\url{https://www.dji.com/it/flysafe/airsense}.}
Moreover, uAvionix is working on small ADS-B transceivers for drones,
which use the commercial frequencies $1090$ MHz and $978$ MHz for ADS-B In
and $1090$ MHz for ADS-B Out.%
\footnote{\url{https://uavionix.com/products/ping1090}.}
However, ADS-B radio stations are not designed
to provide coverage at VLL altitudes \cite{Guterres2017}.
Considering the limitations introduced by interferences
and by the buildings in urban high-density traffic environments,
it remains unclear how the tracking services will be implemented
in UTM infrastructures \cite{Ferreira2021}.

It should be noted that some research groups are
investigating alternative solutions for ADS-B like systems,
aimed at replacing ADS-B for UTM \cite{Lin2019}.
The proposed solutions are based
on different protocols, some of which proprietary ones,
such as FLARM (an acronym based on ``flight alarm''),
other based on wireless standards, such as
4G LTE (Long-Term Evolution), LoRa (Long Range),
APRS (Automatic Packet Reporting System), and others
(see \cite{Minucci2020, Lin2020} and references therein).

The performance analysis of ADS-B
in terms of packet-loss ratio
has been considered in some papers
(see e.g. \cite{Strohmeier2014,Sch2014}).
The analysis in \cite{Strohmeier2014} is based on
ADS-B messages recorded over a 14-day period with a USRP-based receiver.
It shows the impact of several variables,
also related to weather conditions,
on the packet-loss ratio,
%
%
With reference to publicly available
data of the OpenSky network,
in \cite{Sci2019} an analysis
of the overall (i.e., system-related)
packet loss of ADS-B is derived.
In particular, besides providing
an accurate assessment with respect
to the TX-RX distance,
the packet-loss ratio with respect to
the network congestion is evaluated.
The results confirm that such a congestion
is strictly correlated
with the packet-loss ratio.


\section{The proposed architecture}

\begin{figure}[!t]
\centering
\includegraphics[width=0.95\columnwidth]{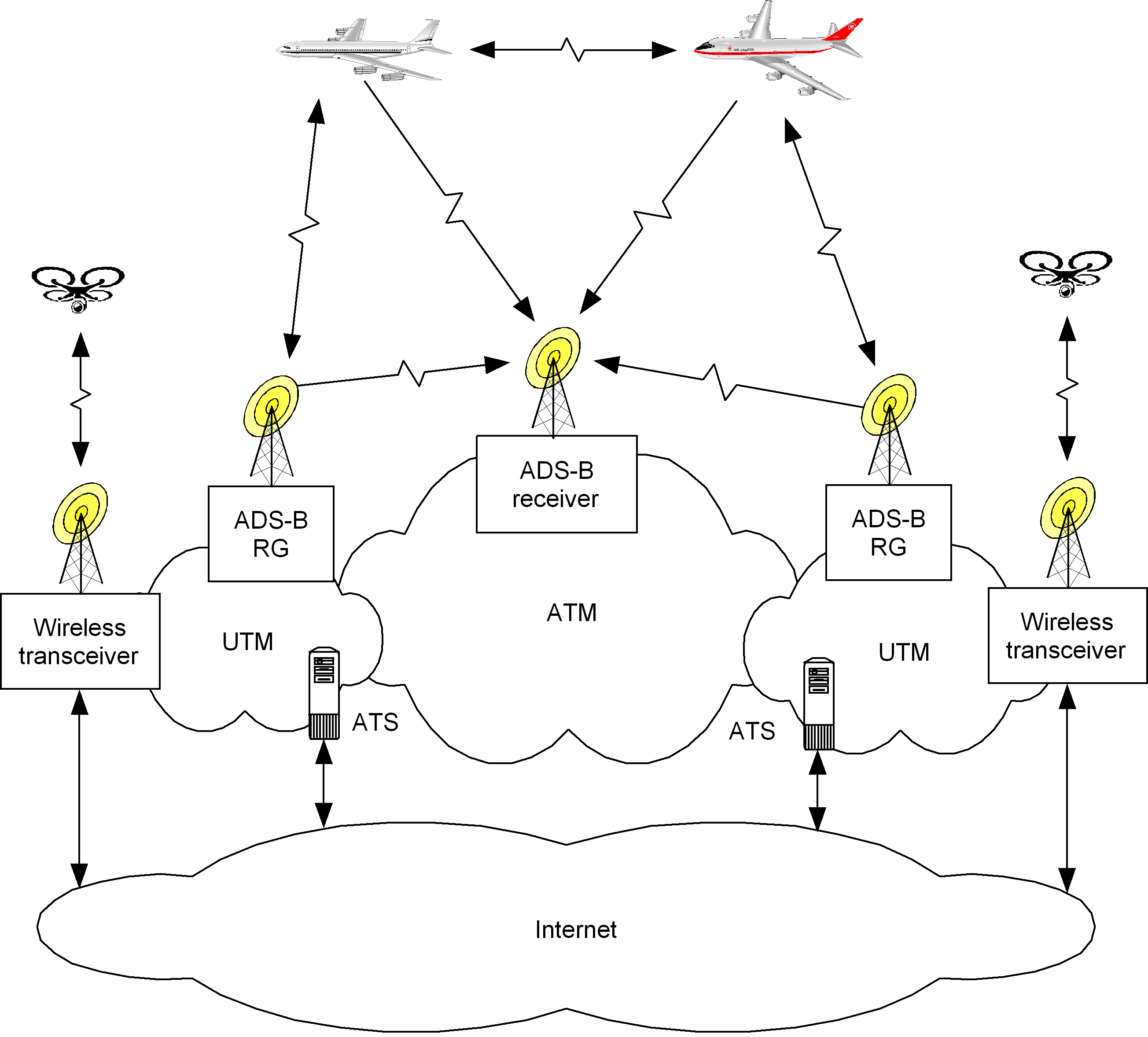}
\caption{The proposed network architecture.}
\label{fig:arch}
\end{figure}

The proposed surveillance solution represents
an \emph{enhanced ADS-B network},
which leverages the existing
ADS-B infrastructure, i.e.,
it can be realized simply by implementing
new features as an add-on to the existing ADS-B network.
The architecture of the
proposed network is depicted in
Fig.~\ref{fig:arch},
and comprises three segments:

\begin{itemize}

\item \emph{Air-segment}:
it is composed by aircrafts and larger drones
employing ADS-B In/Out devices,
as well as smaller drones
connected to the Internet.%
\footnote{Each drone could be directly connected
to the Internet, by means of different radio technologies.
Alternatively, the pilot ground station
can be used as an Internet gateway.}

\item \emph{Ground-segment}:
it consists of existing ADS-B
ground receivers,%
\footnote{The ADS-B protocol also provides
ground transmission nodes, however they are not
used for transmitting positional data
\cite{sun1090mhz}.}%
as well as programmable general-purpose transceivers
to be used as Internet-connected
\emph{ADS-B Radio Gateways} (ADS-B RGs).

\item \emph{Cloud-segment}:
it is based on strongly reliable Internet
\emph{air traffic servers} (ATSs) intended
for authenticating, queuing, processing, and routing
massive amounts of positional data.

\end{itemize}

In the following, we describe in more detail the
functional requirements of the systems and devices
required
by the innovative architecture.

\subsection{Drone on-board system}

Drones not equipped with a traditional ADS-B system
can still retrieve their own positional data
from the on-board situational
awareness subsystems.
After network authentication,
such information is properly encoded,
including an identification string,
and almost-synchronously sent
(nearly at 2 Hz, like in ADS-B)
to the ATSs network (i.e., into the cloud)
using an Internet connection, which can be based
on different radio technologies, such as, e.g., 4G/LTE, 5G, WiFi, or LoRa.
Similarly, drones can retrieve information
through the same Internet connection
about near aircrafts from the ATSs network
to correctly manage the self-separation procedures.

The most efficient way to implement
the communication protocol
between the drones and the ATS is
represented by the
\emph{Request/Reply} paradigm,
due to the synchronous communication needs
and the large number of requests
that the ATS server will need to process \cite{Kou2016}.

\subsection{ADS-B Radio Gateway (ADS-B RG)}
\label{sec:adsb-rg}


The main goal of an ADS-B RG
is to receive from the cloud ATSs network
the positional data of the drones belonging to
its operation area,
format it as ADS-B messages, and
rebroadcast, as an ADS-B Out,
the resulting ADS-B radio messages,
which are used by both ATM and aircrafts
equipped with conventional ADS-B receivers.
Similarly, an ADS-B RG can work as ADS-B In, by
receiving and decoding the ADS-B radio messages coming from
aircrafts under its operation area, and
send the positional data to the ATSs network.

In order to increase
the system capacity,
a \emph{cellular} approach can be used for
allowing resource reuse.
Indeed, each ADS-B-RG is associated
with a limited coverage area
(i.e., a \emph{cell}).
By reducing the cell dimension, capacity
can be efficiently scaled,
by allowing the system to
serve a larger number of drones.


Since ADS-B off-the-shelf devices have not been
designed to change the aircraft ICAO
identifier at run-time.
a reconfigurable transmitter must be developed
to this aim.
The SDR paradigm represents a convenient solution
to implement such ADS-B gateways,
because of its  flexibility features and ease of programming.
Furthermore, SDR allows one to implement an
ADS-B RG by using general-purpose hardware.
In particular, packages for efficient networking
are already available as core modules
of the GNU Radio framework.
Moreover, tools for ADS-B message decoding
have been developed by the GNU Radio community
as \emph{out-of-tree} (OOT) modules%
\footnote{\url{https://github.com/mhostetter/gr-adsb}.}.

Since multiple ADS-B RGs
could be used to broadcast
the same information,
the most efficient way to implement
the communication between ATS nodes and
ADS-B RGs is represented
by the \emph{Publish/Subscribe} paradigm
\cite{Boc2018}.

\subsection{Air Traffic Server (ATS)}

ATS nodes receive from drones and ADS-B RGs
huge amounts of positional data
about drones and aircrafts equipped with
ADS-B transmitters as request messages.
The ATS network must offer
high performance in terms of
reliability, availability, and security
\cite{Mah2015,Has2014}.
The positional datum received by a drone
is encoded following the correct ADS-B
data format \cite{sun1090mhz}
and readily published to be made available
to the ADS-B RGs covering the airspace
near the drone itself.
In the meantime, each ATS node is able to retrieve
air traffic information about the airspace
the drone is covering from both
ADS-B RGs and the global flight tracking services,
and encapsulate them into the reply messages
for enhancing the drone situational awareness.

\section{Performance analysis}


To obtain some simple
analytical results, we consider only the
broadcast information coming from drones
and transmitted by the ADS-B-RGs.
It can be shown that for the reverse path,
i.e., the flow from the aircrafts
equipped with ADS-B transmitters
to the drones, a similar
modeling approach can be used.

The following assumptions and notations
are considered in the analysis:
\textit{(i)} $K$ ADS-B-RGs operate on distinct cells;
\textit{(ii)} the global requests are memoryless with rate $\lambda$;
\textit{(iii)} $P_k$ represents the probability that one request is
received from the cell covered by the $k$-th ADS-B-RG;
\textit{(iv)} all the ADS-B-RGs operate at the deterministic
service rates $\mu_1 = \mu_2 = \ldots = \mu_K = \mu = (1/120) \cdot 10^{6}$ s$^{-1}$
(the duration of an ADS-B message is $120 \, \mu$s);
\textit{(v)} all the service queues are infinite-length.

\begin{figure}
\centering
\includegraphics[width=0.85\columnwidth]{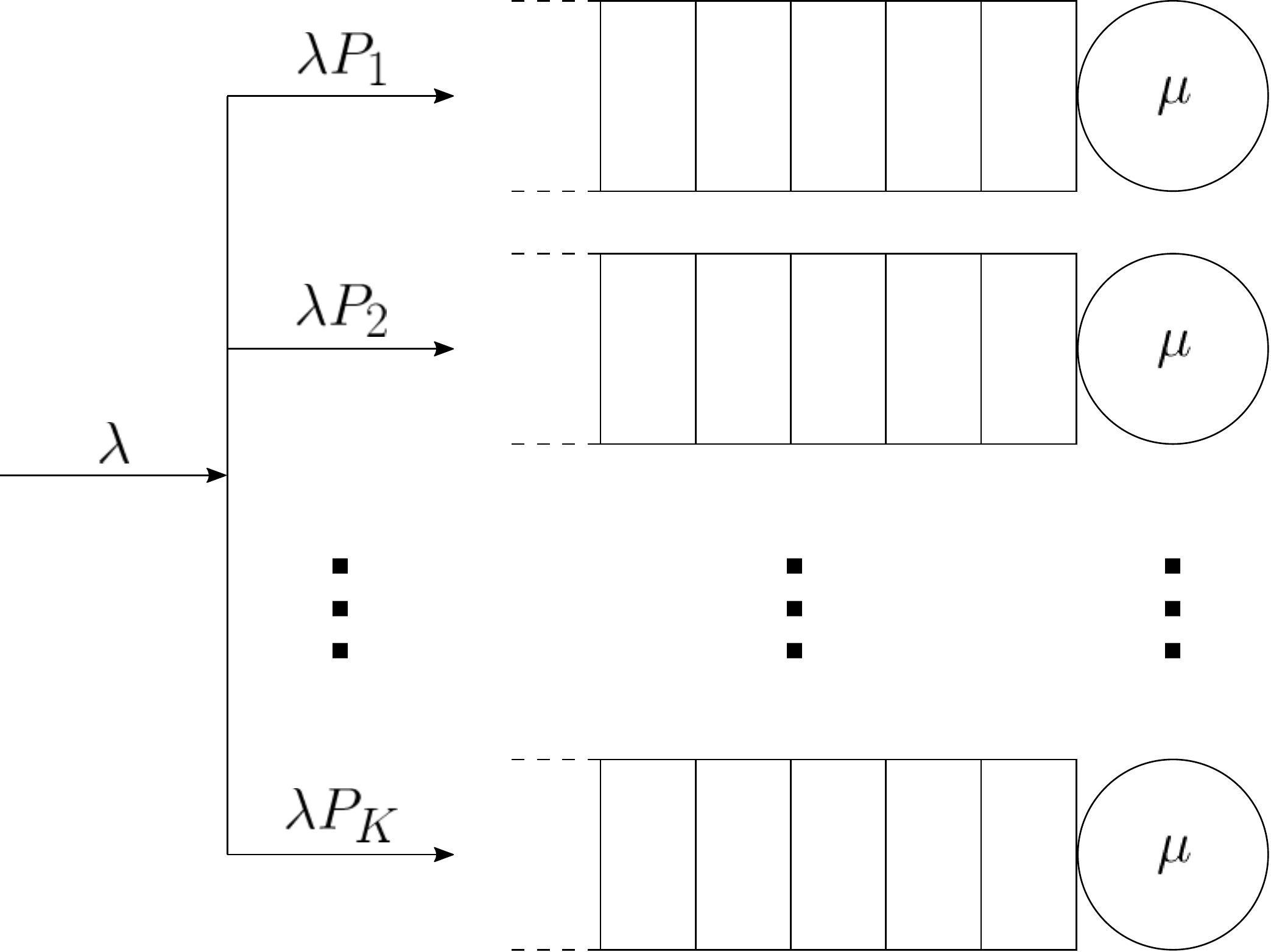}
\caption{Simplified queueing model (parallel M/D/1 queues) for the performance
analysis.}
\label{fig:queue-model}
\end{figure}

We refer to the queueing model
of Fig.~\ref{fig:queue-model},
which represents an architecture
where the network of ATSs
implements the association
between requests and ADS-B-RGs,
on the basis of both coverage and positional 
information,
and manages the ADS-B-RG
queues.
Since the association
depends on the users distribution
within the covered area,
which is not known in advance,
we considered
a model composed by $K$ parallel
M/D/1 queues%
\footnote{Following Kendall notation \cite{Ken1953,Ken2005},
M/D/1 stands for an infinite dimension queue
characterized by memoryless arrivals,
deterministic service rate,
and a single server.},
where the $k$-th queue
is characterized by the arrival rate
$\lambda P_k$ and the service rate $\mu_k = \mu$.

Let $P_{\text{max}} \eqdef \max_{k\in\{1,2,\ldots,K\}} P_k$,
it must be $\lambda P_{\text{max}} < \mu$
to assure system stability.
Since the ATSs use infinite-length queues,
the system blocking probability $P_\text{B}$ will be zero.
The average number $\overline{N}$ of requests
and the average waiting time $\overline{T}$
of requests in the system can be evaluated as
\begin{align}
\overline{N} &= \sum_{k = 1}^{K} \overline{N}_k =
\sum_{k = 1}^{K} \left[\rho_k + \frac{1}{2}\left(\frac{\rho_k^2}{1-\rho_k}\right)\right]
\label{eq:N-ave}\\
\overline{T} &= \sum_{k = 1}^{K} P_k \, \overline{T}_k =
\sum_{k = 1}^{K} P_k \, \left[\frac{1}{\mu} + \frac{\rho_k}{2\mu(1-\rho_k)}\right]
\label{eq:T-ave}
\end{align}
with $\rho_k \eqdef \lambda \, P_k/\mu$.%
\footnote{The average number of requests in the system
can be also obtained from the average waiting time of requests
in the system exploiting the Little's theorem \cite{Ber1992}, i.e.,
$\overline{N} = \lambda \overline{T}$.}
When the drones are uniformly distributed
within the covered area,
i.e., $P_1 = P_2 = \ldots = P_K = 1/K$,
\eqref{eq:N-ave} and \eqref{eq:T-ave}
simplify to
\begin{align}
\overline{N} &= \frac{\lambda}{\mu}
+\frac{\lambda^2}{2\mu(K\mu-\lambda)}
\label{eq:number-avg} \\
\overline{T} &= \frac{1}{\mu}
+\frac{\lambda}{2\mu(K\mu-\lambda)}
\label{eq:time-avg}
\end{align}
with the stability condition becoming $\lambda < K \mu$.

According to the ADS-B protocol,
positional messages are sent every $0.5$ s,
i.e., the transmitting
rate produced by each ADS-B user
is $R_{\text{ADS-B}} = 2$ Hz.
Such a condition allows one to set
an upper-bound for the
\textit{system capacity}:
\begin{equation}
N_{\text{max}} < \frac{K\mu}{R_{\text{ADS-B}}}
\end{equation}
with $N_{\text{max}}$ denoting
the maximum amount of users that
can be served.
As readily seen by \eqref{eq:number-avg} and
\eqref{eq:time-avg},
when $K$ increases a saturation
effect of the performance parameters
is observed, i.e., the average waiting
time boils down to the
time duration $1/\mu$ of a single ADS-B
message.

As stated in \cite{Sci2019}, the packet-loss ratio
is strictly related to the overall system
capacity and congestion:
a higher aircraft congestion entails
a larger number of collisions
on the communication link,
i.e., a higher probability for the transmitters
to choose overlapping time slots
for the transmissions.
The performance analysis reported
above does not consider
possible collisions with ADS-B signals
coming from legacy devices.
In this respect, since ADS-B-RGs are
intended to be implemented
using the SDR paradigm,
\emph{Carrier Sensing Multiple Access} (CSMA)
could be a suitable technique
for managing the packet-loss ratio,
introducing a controlled degradation
in terms of system capacity.
In this case, more complicated approaches
should be used for modeling
the queueing stack and evaluating performances.

\begin{figure}
\centering
\includegraphics[width=1.00\columnwidth]{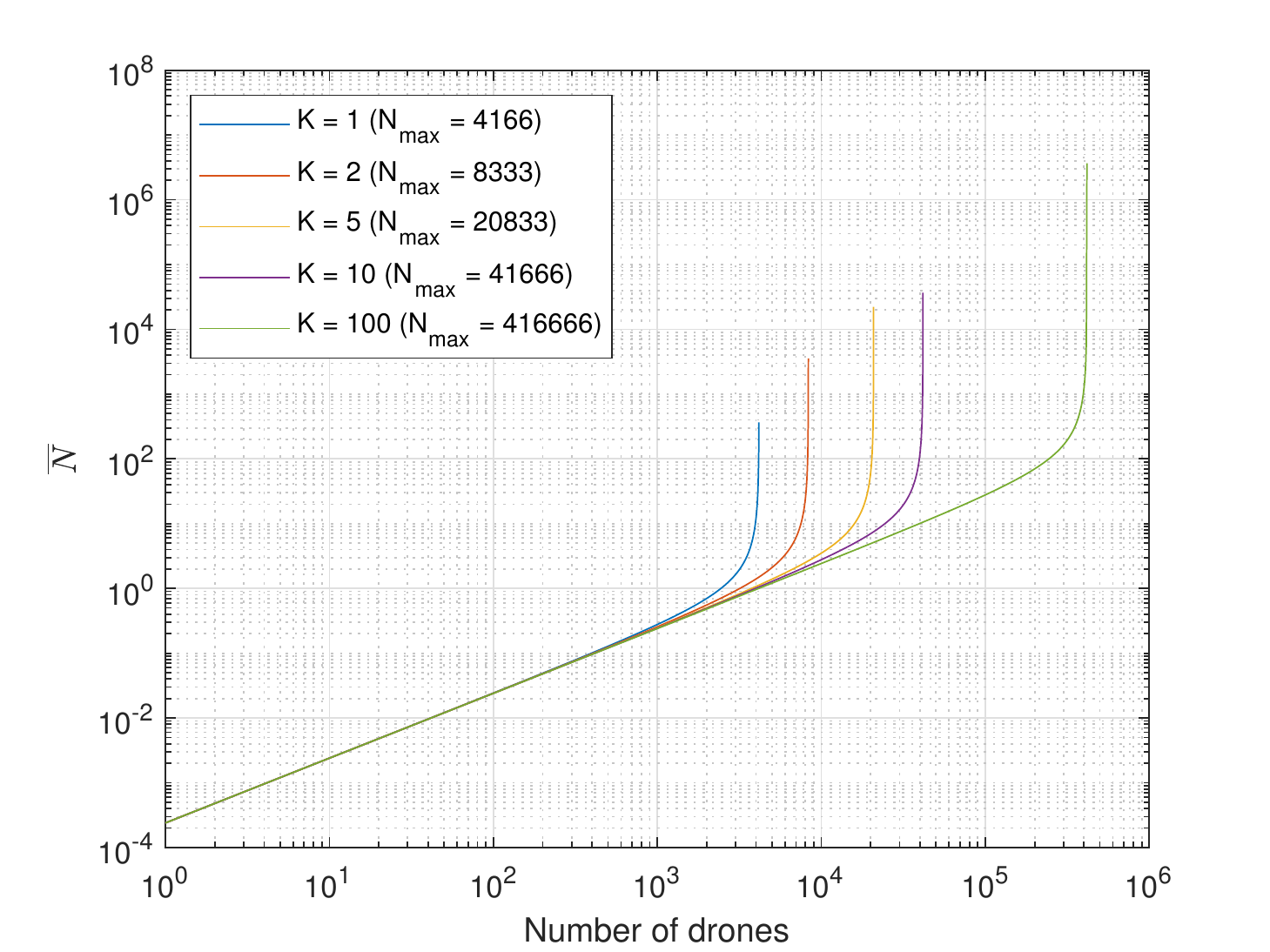}
\caption{Average number of requests in the system as a function of
the number of  served drones.}
\label{fig:queue-N}
\end{figure}

\section{Results}

\subsection{Numerical results}

In this section numerical 
results are provided,
aimed at assessing the system capacity and latency
of the proposed architecture.
The performances have been evaluated
assuming uniformly distributed drones
among the covered area
for different numbers of ADS-B-RGs, i.e.,
$K \in \{1, 2, 5, 10, 100\}$.

\begin{figure}
\centering
\includegraphics[width=1.00\columnwidth]{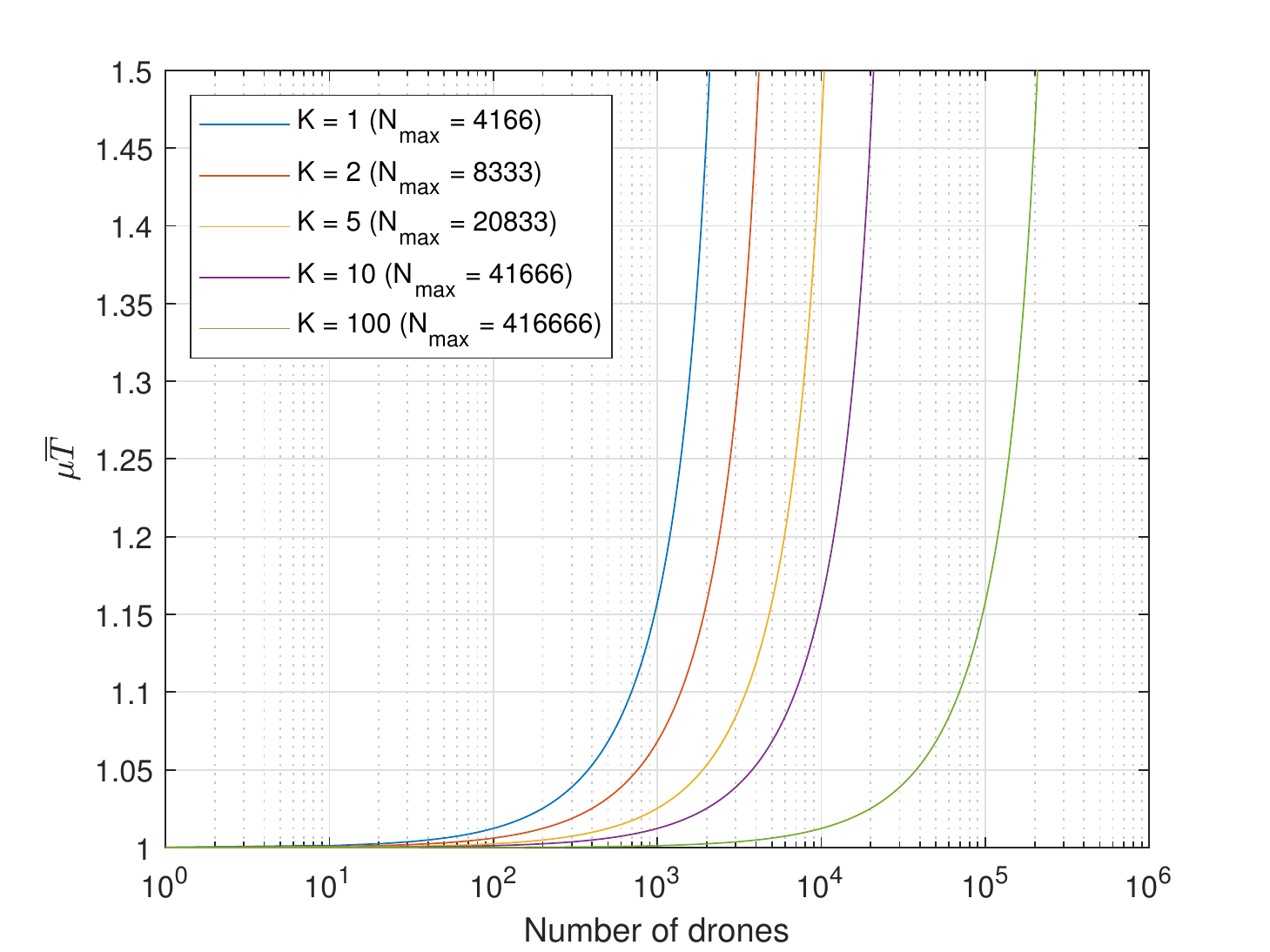}
\caption{Normalized average waiting time of requests in the system as a function of
the number of  served drones.}
\label{fig:queue-T}
\end{figure}

Fig.~\ref{fig:queue-N} shows the average number of requests
in the system as a function of the number of served drones.
It can be readily noted that, although the stability condition
is granted, when the number of drones to be served approaches
the maximum system capacity the number of requests in the system
dramatically increases.
A similar behavior can be observed for the
average waiting time of requests in the system
(normalized to $1/\mu$),
reported in Fig.~\ref{fig:queue-T}.
as a function of the number of served drones.
In particular, this curve shows
that $K$ needs to be properly
designed to accommodate
the latency constraints of the ADS-B standard.

\subsection{Experimental results}

For a first proof-of-concept of the approach,
experimental results have been
focused on the advances
in physical-layer technologies.
Specifically, as stated in section~\ref{sec:adsb-rg},
the TX segment of an ADS-B RG has been implemented
from scratch as an OOT module
into the GNU Radio (v3.8) framework running under
Rocky Linux (v8.5).
In order to test our ADS-B SDR implementation,
a simple C++ application emulating
the message passing between the ATS
and the ADS-B RG was also developed.

The ADS-B RG TX segment gets the asynchronous
ADS-B messages to be transmitted as subscriber
of the ATS publisher using
the \emph{ZeroMQ} networking library%
\footnote{\url{http://zeromq.org}.},
already embedded in GNU Radio.
Each ADS-B message ($112$ bit) is first PPM-binary modulated
at $1$ MBaud.
Then, a short preamble is added,
such that a burst of $120$ $\mu$s
is obtained \cite{sun1090mhz}.
The output burst coming from the aforementioned ADS-B modulator
has to be served at $2$ MHz; if the sample rate of the flowgraph
needs to be set at an higher frequency,
an upsampling procedure
must be provided.

Two functional experiments have been carried out:
the first one considers a ``loopback''
GNU Radio Companion flowgraph,
aimed at validating the correctness of the
modulated burst using
the ADS-B decoding library
provided by the \emph{gr-adsb} module%
\footnote{\url{https://github.com/mhostetter/gr-adsb}.}.
The second experiment implements the entire ADS-B RG
transmitting chain, with the actual RF signal
generated by using
an Ettus USRP E310%
\footnote{\url{https://www.ettus.com/all-products/e310}.}
as RF front-end
and the driver module \emph{gr-iio} provided by
Analog Devices%
\footnote{\url{https://github.com/analogdevicesinc/gr-iio}.}.
The resulting RF signal is demodulated and decoded
using a commercial ADS-B receiver
(Garrecht Avionik GmbH TRX-1090 ADS-B Receiver).

\begin{figure}
\centering
\vspace{5mm}
\includegraphics[width=0.95\columnwidth]{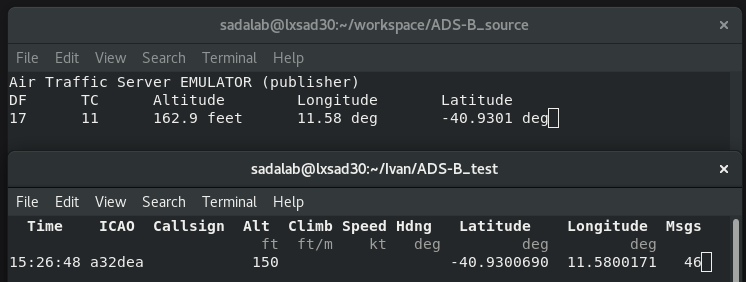}
\caption{A screenshot of GNU Radio loopback experimental results.}
\label{fig:res}
\end{figure}

In both experiments, the parameters
of the emulated/emitted ADS-B signal
are set as DF=17, CA=5, TC=11, ICAO=A32DEA.
Furthermore, the position varies linearly in
altitude, longitude, and latitude,
with respect to time.
Both functional experiments were successful,
i.e., the decoders used for validation
(i.e., gr-adsb and the commercial ADS-B receiver)
were able to correctly decode the emulated/emitted ADS-B signal.
Figure~\ref{fig:res} shows as an example
the emulated published
data from the ATS on the top shell,
and the decoded data on
the bottom shell.


\section{Conclusion}

In this paper we presented a proof-of-concept
of a cloud-assisted network for the integration
of cooperative small UAVs in the
ADS-B surveillance system for unmanned traffic management.
With respect to state-of-the-art solutions belonging to the class of
ADS-B like systems, the proposed solution is not aimed
at replacing ADS-B and does not introduce
a different surveillance protocol.
Instead, it preserves the current ADS-B network,
while integrating and enriching it with cloud-based
features to include in the civil airspace
those flying vehicles that
cannot host ADS-B transceivers.

The proposed solution heavily relies
on cloud-computing infrastructures,
as an extension of the current
ADS-B infrastructure.
Indeed, cloud computing allows efficient processing
of massive amounts of positional data,
solving also some security issues related to
the authentication of managed vehicles,
by means of network
authentication services.
Finally, due to their flexibility and efficient networking features,
the interface between ADS-B network
and cloud infrastructures
is implemented by SDR devices
acting as ADS-B gateways.

In future work,
the functionalities of the cloud segment 
should be implemented in detail
into the simulation setup.
Moreover, an exhaustive evaluation
of the attainable performances
(in terms, e.g., of network capacity, 
message loss, and latency)
should be carried out,
which would take into account
the actual legacy ADS-B data traffic
and the processing delay 
at each network node.



\bibliographystyle{IEEEtran}
\bibliography{ADS-B_bibliography}

\end{document}